# Performance of an Enhanced Afterloader with Electromagnetic Tracking Capabilities for Channel Reconstruction and Error Detection

---

Daline Tho[1,2], Marie-Claude Lavallée[1,2], and Luc Beaulieu[1,2]

[1] Département de radio-oncologie and Centre de recherche du CHU de Québec, CHU de Québec, Quebec, Quebec,

G1R 3S1,Canada

[2] Département de physique, de génie physique et d'optique, et Centre de recherche sur le cancer, Université Laval,

Quebec, Quebec, G1V 0A6, Canada

Corresponding author: Luc Beaulieu

Email: Luc.Beaulieu@phy.ulaval.ca

Mailing: L'Hôtel-Dieu de Québec
4672-6, rue McMahon

Quebec, Quebec, Canada, G1R 3S1

# Performance of an Enhanced Afterloader with Electromagnetic Tracking Capabilities for Channel Reconstruction and Error Detection

Daline Tho[1,2], Marie-Claude Lavallée[1,2], and Luc Beaulieu[1,2]

[1]Département de radio-oncologie et Centre de recherche du CHU de Québec,CHU de Québec,

Université Laval, Québec, Québec, Canada

[2] Département de physique, de génie physique et d'optique et Centre de recherche sur le cancer de

l'Université Laval, CHU de Québec, Québec,Québec,Canada,

October 2020

Short title: Flexitron with Reconstruction Capability

Corresponding author: Luc Beaulieu

email: Luc.Beaulieu@phy.ulaval.ca

## Abstract

**Purpose:** To assess catheter reconstruction and error detection performance of an afterloader (Elekta Brachytherapy, Veenendaal, The Netherlands) equipped with electromagnetic (EM) tracking capabilities. **Materials/Methods:** A Flexitron research unit equipped with a special check-cable in which is integrated an EM sensor (NDI Aurora V3) that enables tracking and reconstruction capability. Reconstructions of a 24 cm long catheter were performed using two methods: continuous fixed-speed check cable backward stepping ( at 1, 2.5, 5, 10, 25 and 50 cm/s) and stepping through each dwell position every 1 mm. The ability of the system to differentiate between two closely located (parallel) catheters was investigated by connecting catheters to the afterloader and moving it from its axis with an increment of 1 mm. A robotic arm (Meca500, Mecademic, Montreal) with an accuracy of 0.01 mm was used to move the catheter between each reconstruction. Reconstructions were obtained with a locally weighted scatter-plot smoothing algorithm. To quantify the reconstruction accuracy, distances between two catheters were computed along the reconstruction track with a 5 mm step. Reconstructions of curve catheter paths were assessed through parallel and perpendicular phantom configuration to the EM field generator. Indexer length and lateral error were simulated and ROC analysis was made. **Results:** Using a 50 cm/s check cable speed does not allow accurate reconstruction. A slower check cable speed results in better reconstruction performance and smaller standard deviations. At 1 cm/s, a catheter can be shifted laterally down to a 1 mm and all paths can be uniquely identified. The optimum operating distance from the field generator (50 to 300 mm) resulted in a lower absolute mean deviation from the expected value (0.2±0.1mm) versus being positioned on the edge of the electromagnetic sensitive detection volume (0.6±0.3mm). Reconstructions of curved and circular paths with a check cable speed under 5 cm/s gave a 0.8 mm±0.5 mm error, or better. All indexer and lateral shifts of 1 mm were detected with a check cable speed of 2.5 cm/s or under.

**Conclusions:** The EM-equipped Flexitron afterloader is able to track and reconstruct catheters with high accuracy. A speed under 5 cm/s is recommended for straight and curved catheter reconstructions. It allows catheter identification down to 1 mm inter-catheter distance shift. The check cable can also be used to detect common shift errors.

***Keywords*** — Brachytherapy, Afterloader, Electromagnetic Tracking, Reconstruction, Catheter

# 1   Introduction

Brachytherapy is used to treat numerous body sites including prostate, breast and cervix. Sparing the healthy surrounding tissues and achieving an optimal dose distribution escalation are the procedure's main advantages. It can be used as a monotherapy or as a boost combined with other treatments. A variety of dose-fractionation schedules are possible, with the option of increasing the dosage as high as 19 Gy per fraction [1]. During a high dose rate (HDR) brachytherapy treatment, positioning and timing accuracy are of critical importance because it delivers large doses of radiation in a reduced number of visits. Therefore, not knowing the source's trajectory in the catheter with the best accuracy and precision can lead to underdosage of the target or an overdosage of the organs at risks. Accurate registration of a catheter/applicator to the anatomy plays an integral role in the uncertainty of the delivered dose and how it is distributed [2,3].

Many papers have studied the integration of an electromagnetic (EM) system in a treatment verification process and its impact when used in a clinical environment [4–9]. The Aurora electromagnetic tracking (EMT) system (NDI, Ontario, Canada) is commonly used in these studies. The EMT's field generator creates an inhomogeneous electromagnetic field within a cubic volume of 50 cm per side. This volume allows a sensor to be placed and tracked. It is proved useful for reconstructing catheters and applicators with a good precision and accuracy (±0.8 mm) [10–12]. Potential treatment errors can be caught by an EMT system before dose delivery [6]. The integration of such a system to an afterloader has been proposed for pre-treatment verification [4]. The use of a hybrid afterloader has also been studied for error detection in multi-catheter interstitial brachytherapy [13]. All those studies using this system are part of a clinical trial where the reported errors and/or uncertainties come from a combination of the setup, the patient motion (e.g. breathing) and the intrinsic system performance itself.

This work explored the performance of the catheter/applicator reconstruction capabilities of an afterloader equipped with an EMT sensor into the check cable in well-controlled conditions. Its objective is to determine the intrinsic performance of the hybrid system using well-known reference points i.e. establishing the baseline performance for this proposed hybrid system.

Comparison between two methods for detection of errors while using the check cable was also studied and analyzed for the system's performance.

## 2 Materials and Methods

### 2.1 Afterloader unit

A prototype afterloader research unit based on the Flexitron system (Elekta Brachytherapy, Veenendaal, The Netherlands) equipped with a special check-cable in which is integrated an EM sensor (NDI Aurora V3, Ontario, Canada), that allows automated tracking and reconstruction capability. This unit is non-clinical and contains a dummy source. The EM sensor is a cylinder wide of 0.8 mm in diameter and long of 4 mm. It has an intrinsic positional reading reproductibility (average standard deviation of data recording of 10 s with the sensor at a fixed position) of $\pm 0.08$ mm.

### 2.2 Catheters reconstruction at different check cable speeds

Catheter's reconstructions were performed using two methods: continuous motion and step-and-record (0 cm/s). The first one consists of a continuous check cable fixed-speed backward motion (retraction). Catheters were reconstructed using different check cable speeds (1, 2.5, 5, 10, 25 and 50 cm/s). The second records steps through specific dwell positions for 10 s (each 1 mm or 5 mm) and therefore bypasses any potential extra signal due to a moving object in an EM field. It will serve as a reference throughout this manuscript.

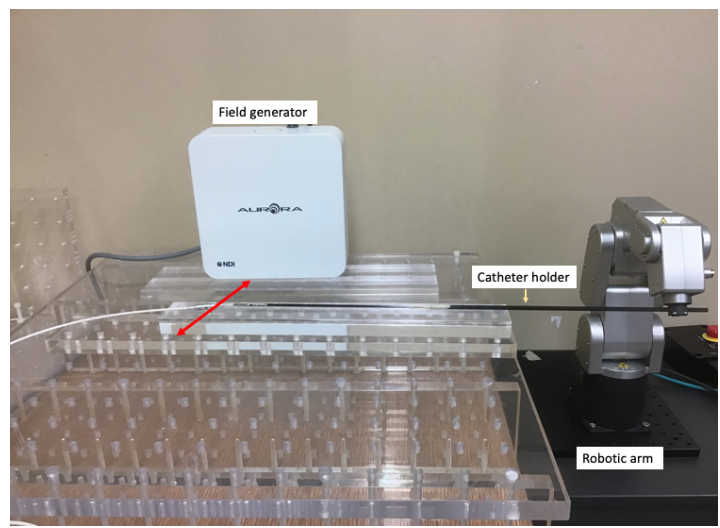

*Figure 1* – Catheter reconstruction experimental setup, here with the catheters parallel to the field generator and the Meca500 robotic arm moving one catheter away, in a perpendicular direction (the red arrow).

### 2.2.1 Catheter translation

A catheter was connected to the afterloader and moved perpendicularly from its original position with an increment of 1 mm. The reconstructions were 24 cm long. A Meca500 robotic arm (Mecademic, Montreal, Canada) (Fig. 1) with 0.01 mm accuracy was used to move catheters between each reconstruction. Reconstructions using step-and-record methods were performed with 1 mm step. Reconstructions were obtained with a locally weighted scatterplot smoothing algorithm like the one used in Poulin *et al.* [12, 14]. The reconstruction distances between two parallel catheters (which is fixed and given by the robotic arm displacements) were computed along the reconstruction track at every 5 mm in order to quantify the reconstruction accuracy. Continuous reconstructions for all check cable speeds were performed and compared to the step-and-record reconstruction (0 cm/s) method used as the reference in this part. The reconstructions were made at 100 mm and 300 mm from the field generator.

### 2.2.2 Catheter's curved trajectories

Reconstruction in a phantom was used to assess the feasibility of a curved catheter reconstruction in two configurations: parallel and perpendicular to the field generator. Five dulerin plastic cylinders with diameters ranging from 22 mm to 56 mm (to mimic the curvature of the source path in clinical applicators) (Fig. 2) were used to reconstruct the circular paths. A smaller tube was used instead of a regular catheter to lower the radial fluctuations. The plastic tube of 1.06±0.01 mm (part number 148-0053, Nordson Medical, Ohio, USA. ) diameter enrolled around each cylinder was reconstructed using the EM-enabled check-cable, again using the step-and-record and continuous approaches. The computed expected circular path takes into account the inner diameter and the wall thickness of the plastic tube used. Adhesive putty was placed around the tube to make sure that it didn't move during all the reconstruction procedure.

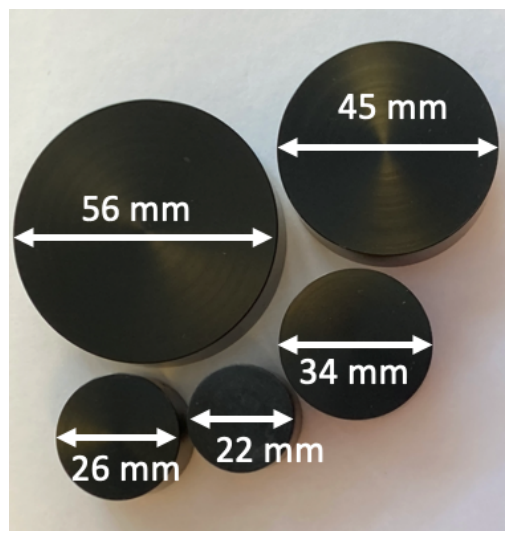

*Figure 2* – Dulerin plastic cylinders with variable dimensions used for circular path reconstruction.

All reconstruction data were compared to the expected circular path. This expected path coordinates were generated with the diameter and the center of the each reconstruction. The radial errors (R) and the vertical errors (Z) were computed separately. Only data along circular path section were used for comparison.

## 2.3    Comparison of clinical ring applicator reconstruction

Reconstruction with an interstitial ring CT/MR applicator set (Elekta Brachy, Veenendaal, The Netherlands) was also performed. All reconstructions were compared to the manufacturer measured source path and the clinical source path at different speeds. The clinical path was extracted by placing the applicator on an AP/LAT film and moving the source with a step size of 5 mm. A Philips BV Pulsera mobile C-arm was used. The rings used had diameters of 26 mm (Part number 110.330A02), 30 mm (110.331A02) and 34 mm (110.332A02), all part of the interstitial ring CT/MR applicator from Elekta.

## 2.4    Error detection

Lateral and indexer (longitudinal) shifts (0.3, 0.5, 1.0, 3.0 and 5.0 mm) were simulated with the robotic arm. Reconstruction of catheter at different speeds were used to define dwell positions and to detect the known introduced errors using two different methods: mean deviation and individual dwell comparison [13, 15, 16]. In the first method, the mean deviation of all dwell positions relative to their reference positions in a catheter is computed and compared to a threshold. If the mean value is smaller than the threshold, the catheter is marked as unshifted. This threshold was set to the intrinsic variability of the Aurora system at $\pm 0.1$ mm [17]. Larger values will be assigned to a shift or motion. In the second approach, the difference between each position and the reference is again extracted. However, instead of computing a mean deviation, a majority rule approach is applied. This means that when a shifted channel path is detected, at least 50% of individual dwell are above the set threshold ($\pm 0.1$ mm) [16].

# 3    Results

## 3.1    Catheter translation

Figure 3 illustrates four reconstruction scenarios. Figure 3A-B shows reconstruction at 50 cm/s far and near the field generator. Panel C shows reconstructions of straight catheter (laterally shifted) at a speed of 1 cm/s. The average error was $0.04 \pm 0.02$ mm. Figure 3D shows reconstructions of curved catheter in a phantom at 10 cm/s, 25 cm/s, and 50 cm/s. Reconstruction of a curved and circular paths with a check cable speed under 5 cm/s gives a 0.8 mm$\pm 0.5$ mm error. A slower check cable speed further resulted in a smaller standard deviation (Fig. 4).

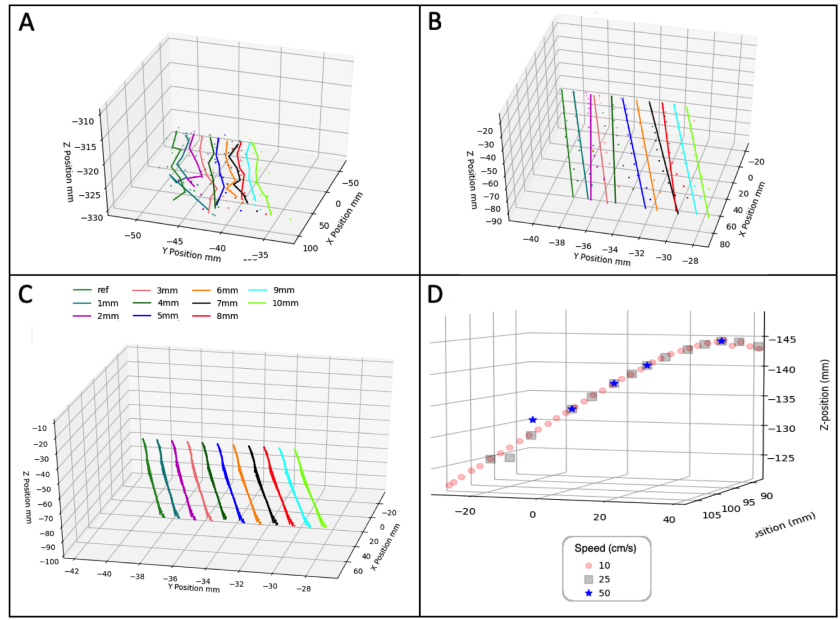

*Figure 3* – Different reconstruction scenarios. A) 50 cm/s at 300 mm from the field generator, B) 50 cm/s at 100 mm from the field generator, C) 1 cm/s at 100 mm from the field generator and D) curved catheters in phantom at 100 mm from the field generator. Each continuous line is a reconstruction from a locally weighted scatter plot. A scatter plot is a raw reconstructions of recorded data.

Figure 4 shows the difference between the continuous motion and the step-and-record methods. Mean differences at all speeds are less than 1 mm when the reconstruction occurs near the generator (under 150 mm from the field generator). On the opposite hand, when the reconstructions are further away from the generator, mean differences are higher between the two reconstruction methods.

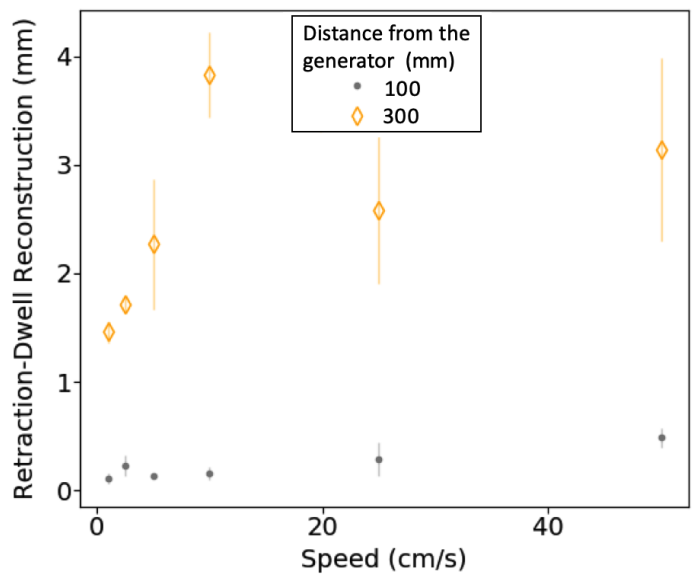

*Figure 4* – Difference between continuous motion and step-and-record reconstruction method of a catheter lateral shift. Two sets of data were obtained: one near the field generator (100 mm from the generator) and far from it (300 mm from the generator).

## 3.2 Catheter's curved trajectories

For every circular path diameter, speed of 25 cm/s and 50 cm/s gave higher mean errors. For all five diameters, R-errors were smaller than Z-errors, with a larger spread of the R-errors distribution (this is shown in figure S-1-5, refer to supplementary material). Reconstructions at 25 cm/s and 50 cm/s showed errors greater than 1.5 mm which was not seen when the speed was at 10 cm/s or lower.

*Table 1* – Paired t-test and p-values to compare speed reconstructions for all curved and circular paths.

| Speed (cm/s) | p-value | |
| :---: | :---: | :---: |
| | Radial (R) errors | Vertical (Z) errors |
| 1 | 0.95 | 0.88 |
| 2.5 | 0.93 | 0.70 |
| 5 | 0.64 | 0.31 |
| 10 | 0.94 | 0.47 |
| 25 | 0.02 | 0.01 |
| 50 | 0.00 | 0.01 |

Absolute mean errors for all speed were compared to the step-and-record reconstruction. A t-test showed that there was not a significant difference between the continuous retraction reconstruction and the step-and-record reconstruction for both R and Z, whether it was a speed of 10 cm/s or under (Table 1). The vertical absolute mean error for all speeds was 0.4±0.3 mm and the radial absolute mean error was 0.3± 0.3 mm. All mean absolute errors of R and Z are shown in figure 5. In circular path reconstruction, mean absolute errors and standard deviations increased with the check cable speed for every diameter. As for radial error, standard deviations were smaller for parallel-to-field-generator reconstructions except for the 22 mm diameter. Vertical mean absolute errors of 0.4±0.3 mm and 0.3±0.2 mm were found for both perpendicular and parallel orientation.

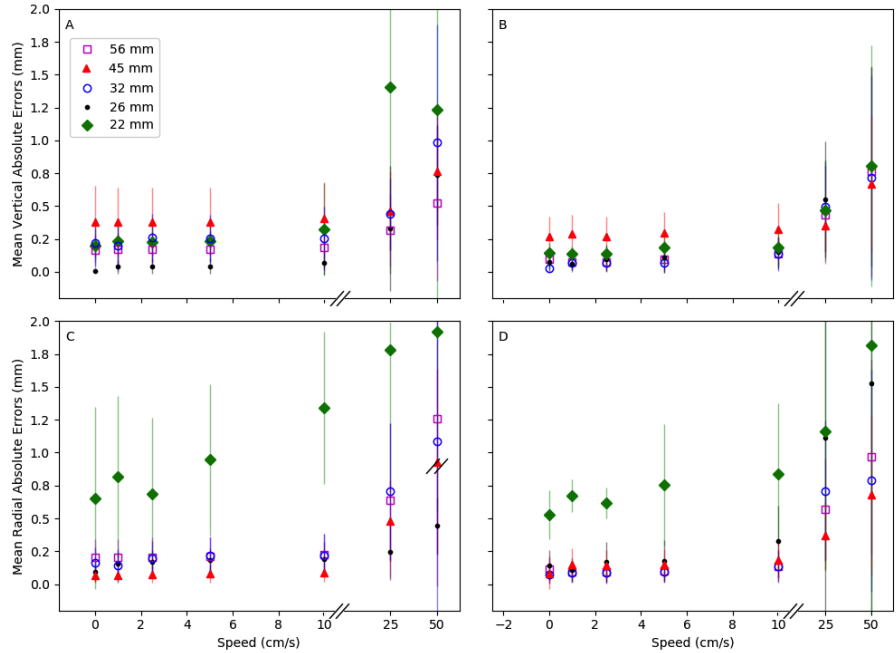

*Figure 5* – Mean absolute R and Z errors for five different circular path diameter reconstructions. A and C are reconstructions parallel to the field generator; B and D are perpendicular to the field generator. The reconstruction at 0 cm/s is the step-and-record method.

## 3.3 Comparison of ring applicator reconstruction

Figure 6 shows vertical and radial mean differences between the manufacturer reconstruction and the check cable reconstruction. Mean increases as the reconstruction speed increases. The same tendency can be noticed between the check cable reconstructions and the clinical source reconstruction. Reconstructions with speeds under 10 cm/s have radial and vertical mean differences under 1.9 mm from the clinical extracted path and under 1.4 mm from the manufacturer path (Fig.6 ). The step-and-record has a maximum mean difference of 0.5±0.2 mm when compared to the factory data. Comparing with the clinical data a mean difference of 0.9±0.3 mm was observed. Difference between the clinical source reconstruction data and the manufacturer reconstruction were all under 1.7 mm. The mean vertical difference was of 0.6±0.5 mm and the mean radial difference was of

0.2±0.2 mm.

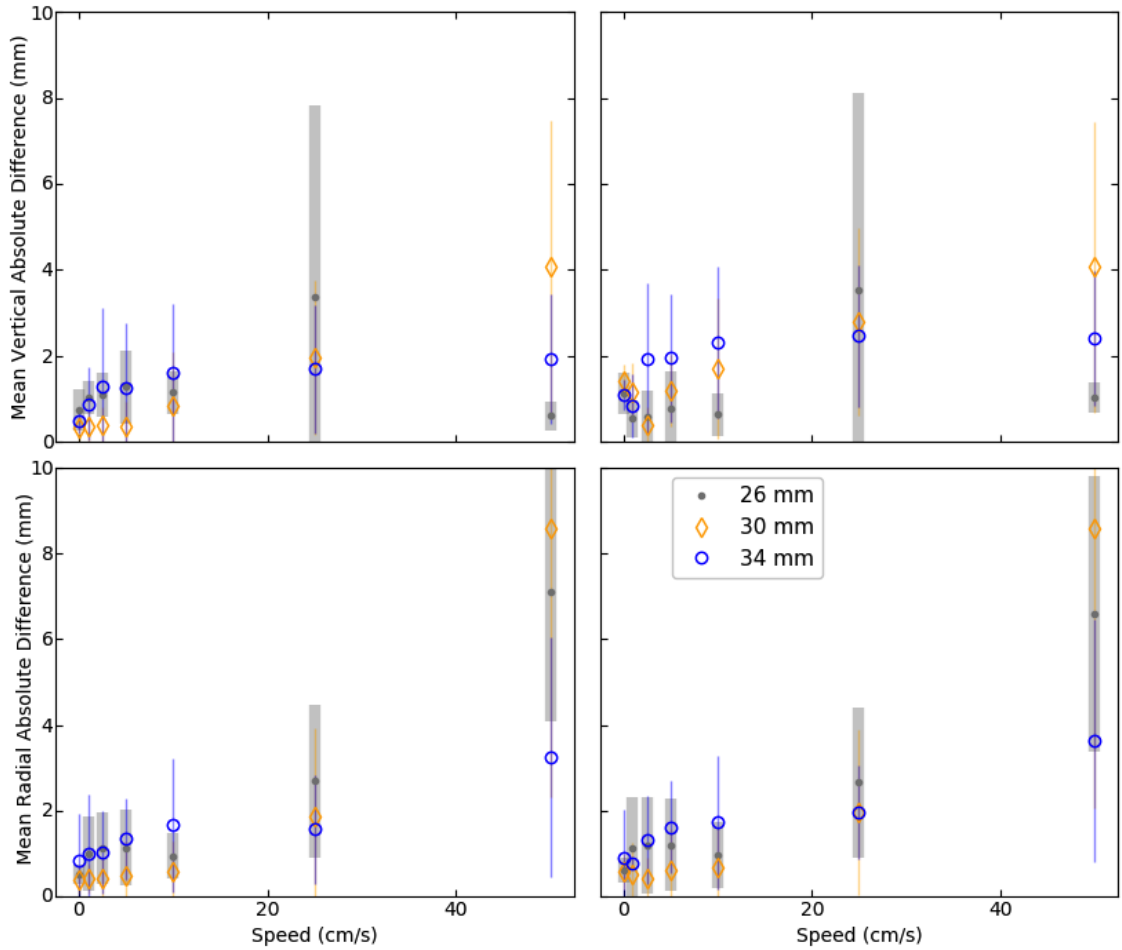

*Figure 6* – The differences between the ring reconstruction using the check cable, the manufacturer measured source path (left panels) and the clinically extracted path (right panels) as a function of the check cable speed.

## 3.4 Error detection

Figure 7 shows the area under the curve (AUC) of a receiver-operator curve (ROC) for different simulated errors. Under 5 cm/s check cable speed, all 1 mm shifts were detected for both longitudinal and lateral errors. For 5 mm shift, all errors were detected with a check cable speed of 25 cm/s and under. At speed of 5 and 10 cm/s, all indexer errors of 5 mm and lateral errors of 3 mm were detected.

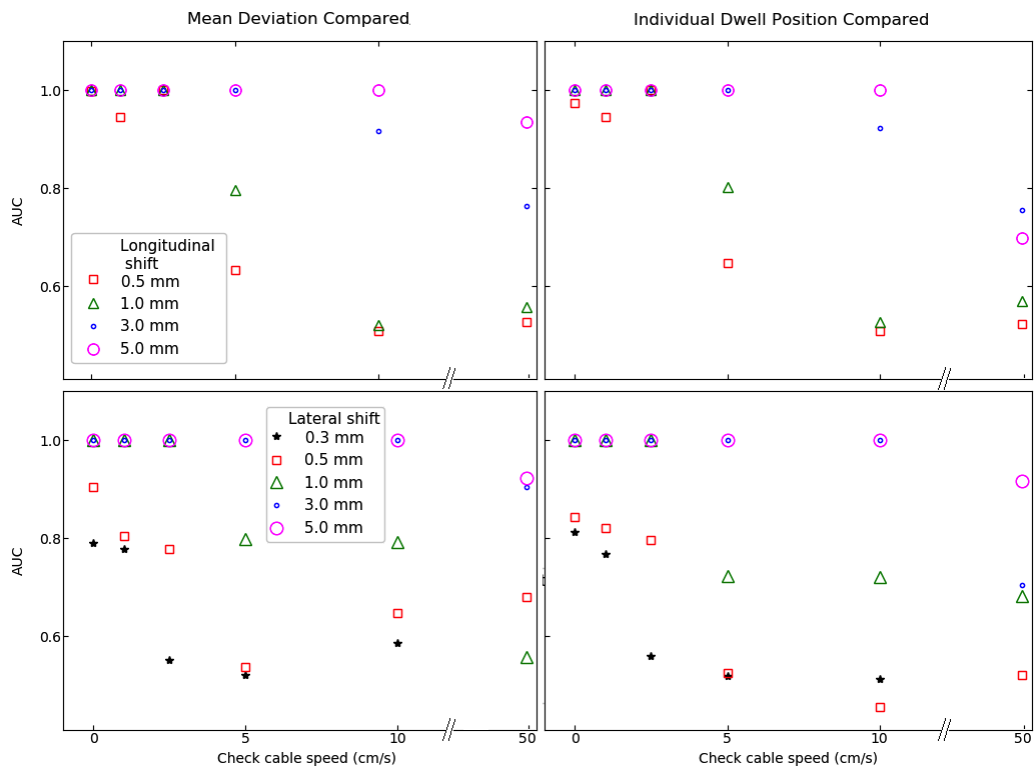

*Figure 7* – Area under a receiver operator curve as a function of the check cable speed for known indexer's length (upper panels) and lateral (lower panels) shifts. Both error criteria were tested: the right panels use the mean deviation while the left panels use individual dwell positions with a threshold at 50 % as presented in the method section.

## 4 Discussion

It was possible to reconstruct a shifted catheter down to a 1 mm with a check cable speed of 1 cm/s. Furthermore, the system was able to identify (uniquely reconstruct) two catheters side-by-side (2 mm apart center-to-center distance). Comparison of the two reconstruction methods (continuous motion and step-and-record) showed their equivalence whether under 150 mm from the generator. However, biased computed coordinates can arise at high check cable speed because of a decrease of sample points and of changes in induced current from the sensor's movement in the electromagnetic field (in particular for circular motions).

The usage of a 50 cm/s check cable speed did not allow accurate reconstruction (Fig. 3A-B). For the reconstruction at greater speed (25 cm/s and 50 cm/s), there were less data points over the whole catheter length which lead to higher uncertainties in comparison to the lower speed. For example, in the smallest cylinder diameter reconstruction, less than 10 reconstruction points were present for a check cable speed of 50 cm/s. The optimum operating distance from the field generator resulted in a lower absolute mean deviation from the expected value (0.2±0.1 mm) rather than being closer to the edge of the electromagnetic sensitive detection volume (0.6±0.3 mm) [10] (Fig.4). A speed of 5 cm/s had a mean absolute deviation of $0.27 \pm 0.08$ mm. It provided a good compromise between reconstruction accuracy and speed.

A fast EMT-enabled check-cable translation is not suitable for channel reconstruction, especially for curved trajectories as there are not enough data recorded for accurate reconstruction, especially while using a LOESS algorithms or the like [4, 12]. However, we have demonstrated that circular path reconstruction is feasible and that there is no significant difference between the continuous motion and the step-and-record method, both at a speed below 10 cm/s (Table 1).

It should be noted that our results suggest that the reconstruction of a circular path appears to have less out-of-plane variability in the parallel-to-field-generator configuration (0.3±0.2 mm) compared to the perpendicular one (0.4±0.3 mm). The larger radial errors for the 22 mm diameter can be explained by the fact that the tube used was too soft. When the check cable passed into the tube, the latter extended and changed in shape (this can be seen in figure S-6-7, refer to supplementary material). To avoid the deformation, we tried to add material (adhesive putty) around the tube, but an obstruction occurred. The check cable does not allow reconstruction of a 22 mm diameter if the tube is too tight.

Motions in the circular path changed the vertical distance between the sensor and the field generator which induced supplementary uncertainties. In a clinical setting, a variety of configuration between these two extremes are likely. Thus, adoption of the lower range of retraction speed should be favored. Mean vertical and radial differences between the manufacturer ring reconstruction data and the clinical source reconstruction data are of the same magnitude. The same goes for the difference between each of them and the extracted reconstruction from the check cable. The difference between the clinical source and the manufacturer data were within the variability of a check cable reconstruction at 1 or 2.5 cm/s (0.4±0.1 mm for curved trajectories).

Results from the error detection study demonstrated that both methods (mean deviation and individual dwell comparison as described in section 2.4) can be used with similar behavior. For example, a 5 mm shift mimicked the error of connecting to a catheter of an adjacent hole in a regular prostate HDR template. This type of error was detected using a check cable speed of 10 cm/s. According to the ROC analysis, any deviation of 1 mm is only reliably detected with a check cable speed of 2.5 cm/s or less in these well controlled conditions. Masitho *et al.* showed that 97 % of shifts over 1.1 mm were detected when reconstruction of catheter were performed with a step-and-record method [16].

The integration of an EM tracking into an afterloader opens up the possibility of automating many tasks once the afterloader is connected to the patient. Thus, it is now possible to detect simple (detecting connection errors) and complex (validating or performing the reconstruction, detecting catheter shifts) issues relatively fast. Most importantly, those verification procedures can be integrated in the treatment workflow (rather than additional manual tasks). The check cable of the hybrid afterloader can be set to a specific speed which is an advantage compared to manual reconstruction.

In this work, the mean absolute deviations computed at 100 mm and 300 mm from the generator are in agreement with the ones found in Damato *et al.* and Poulin *et al.* which are 0.6 ± 0.2 mm and 0.21 mm respectively [6, 12]. It is to be noted that all those reconstructions were performed manually, therefore there were not set at a specific speed. At the edge of the detection volume, Zhou *et al.* recorded sensor positions errors up to 16 mm [8]. This can be noticed in the position difference when comparing reconstructions at 300 mm and 150 mm from the generator.

## 4.1 Limitations

The check cable's diameter used in this paper was smaller than the catheter's inner diameter. This could explain some of the observed radial fluctuations. The objective of the work done on circular paths was to reflect the ideal case for ring applicators and curved trajectories. As such, a smaller tube was used instead of a regular catheter to lower the radial fluctuation. This explains the larger discrepancies with a real applicator geometries, as the source is smaller than the internal applicator's diameter. It should be kept in mind that radial fluctuation can increase because of different applicator's inner diameters.

In this study, check cable speed in-between 5 cm/s and 10 cm/s were not tested because those were not available in the hybrid system. However, for most applications, 5 cm/s and 10 cm/s speeds were found equivalent except for the circular path reconstruction, where 5 cm/s (or lower) led to better results. At this point in time there is no link to an imaging system (and the treated geometry). Error detection is limited to transfer tube swaps and shifts, either along the insertion axis or the perpendicular direction. The latter is true only if one or a few catheters are shifted relatively to the overall reference catheter geometry. If all catheters are shifted in a similar length, the EM-enabled check-cable would not detect a problem without an external absolute positional reference (either from imaging or a reference sensor, ...).

## 5 Conclusion

An evaluation of an EMT-equipped Flexitron afterloader reconstruction performance was done. This work showed that the field generator should be placed parallel to the insertion path as much as possible to reduce out-of-plane variability. Errors in the reconstruction are greater when using a higher check-cable speed. A speed under 10cm/s is recommended for catheter reconstruction. Although for ring applicators and other circular paths, a speed of 5 cm/s or slower should be considered. A comparison with clinical ring applicator was also made. The difference between the clinical source and the manufacturer data were within the variability of a check cable reconstruction at a speed of 1 or 2.5 cm/s. The error detection part led to different speed limits depending on the error's magnitude one would want to detect. All check cable speed of 10 cm/s or under could detect all 5 mm shifts. For a 1 mm shifts to be detected, the check cable speed had to be of 2.5 cm/s or under.

## Acknowledgments

This work was supported by the National Sciences and Engineering Research Council of Canada (NSERC) via the NSERC-Elekta Industrial Research Chair. We thank Edwin Nijenhuis, Eise Jan Kromhout van der Meer, and Bianca Servant for assisting with the research Flexitron afterloader unit.

# Figures and tables

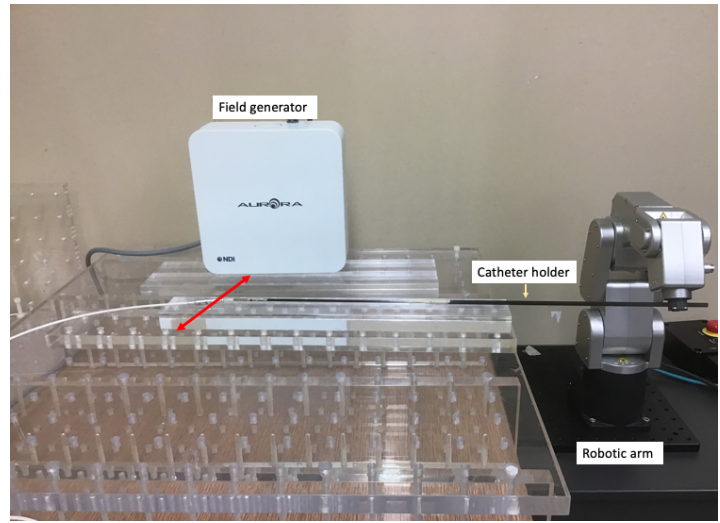

*Figure 1* – Catheter reconstruction experimental setup, here with the catheters parallel to the field generator and the Meca500 robotic arm moving one catheter away, in a perpendicular direction (the red arrow). (repeated from page 3)

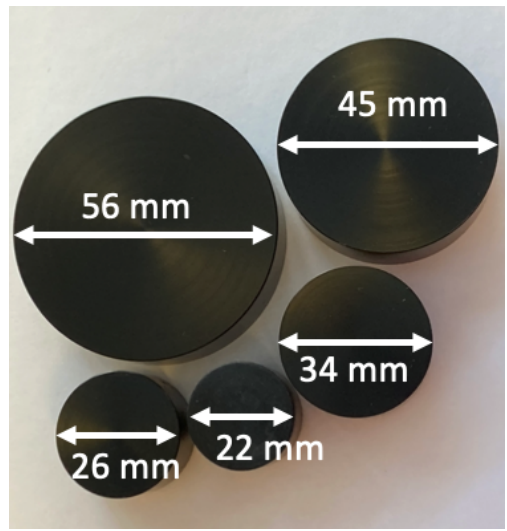

*Figure 2* – Dulerin plastic cylinders with variable dimensions used for circular path reconstruction. (repeated from page 4)

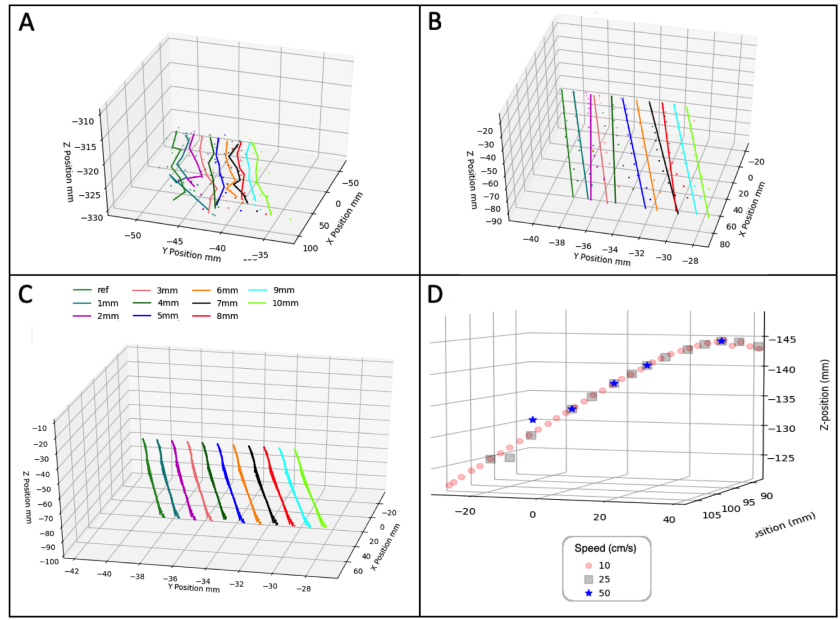

*Figure 3* – Different reconstruction scenarios. A) 50 cm/s at 300 mm from the field generator, B) 50 cm/s at 100 mm from the field generator, C) 1 cm/s at 100 mm from the field generator and D) curved catheters in phantom at 100 mm from the field generator. Each continuous line is a reconstruction from a locally weighted scatter plot. A scatter plot is a raw reconstructions of recorded data. (repeated from page 6)

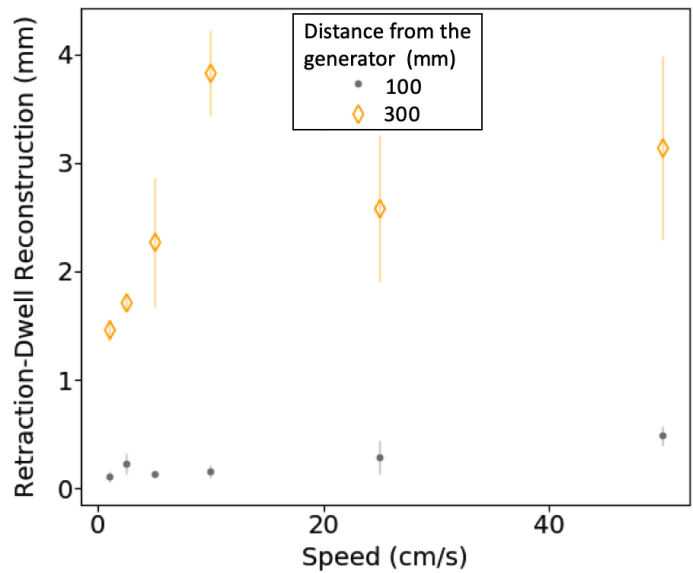

*Figure 4* – Difference between continuous motion and step-and-record reconstruction method of a catheter lateral shift. Two sets of data were obtained: one near the field generator (100 mm from the generator) and far from it (300 mm from the generator). (repeated from page 7)

*Table 2* – Paired t-test and p-values to compare speed reconstructions for all curved and circular paths. (repeated from page 7)

| Speed (cm/s) | p-value | |
|---|---|---|
| | Radial (R) errors | Vertical (Z) errors |
| 1 | 0.95 | 0.88 |
| 2.5 | 0.93 | 0.70 |
| 5 | 0.64 | 0.31 |
| 10 | 0.94 | 0.47 |
| 25 | 0.02 | 0.01 |
| 50 | 0.00 | 0.01 |

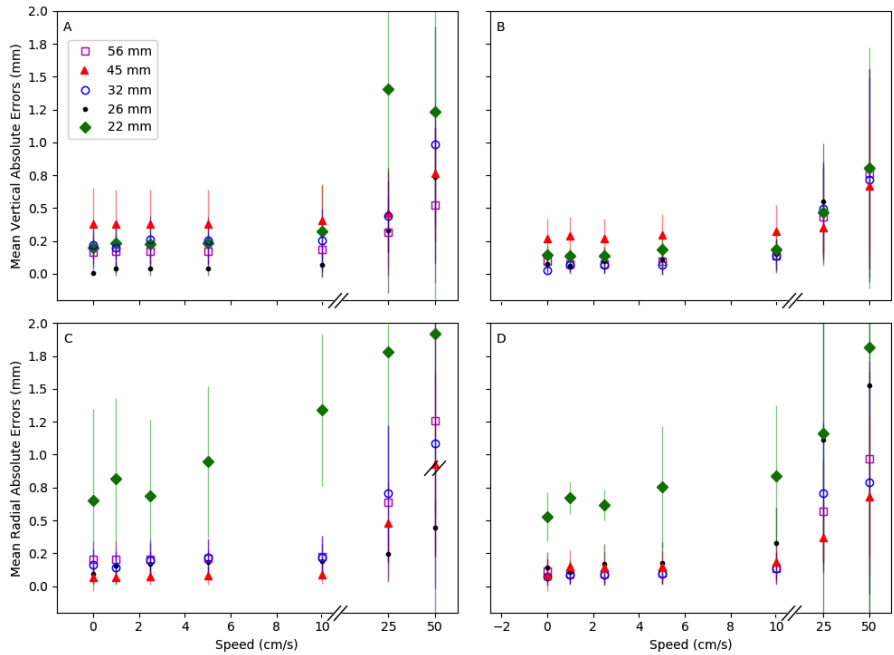

*Figure 5* – Mean absolute R and Z errors for five different circular path diameter reconstructions. A and C are reconstructions parallel to the field generator; B and D are perpendicular to the field generator. The reconstruction at 0 cm/s is the step-and-record method. (repeated from page 8)

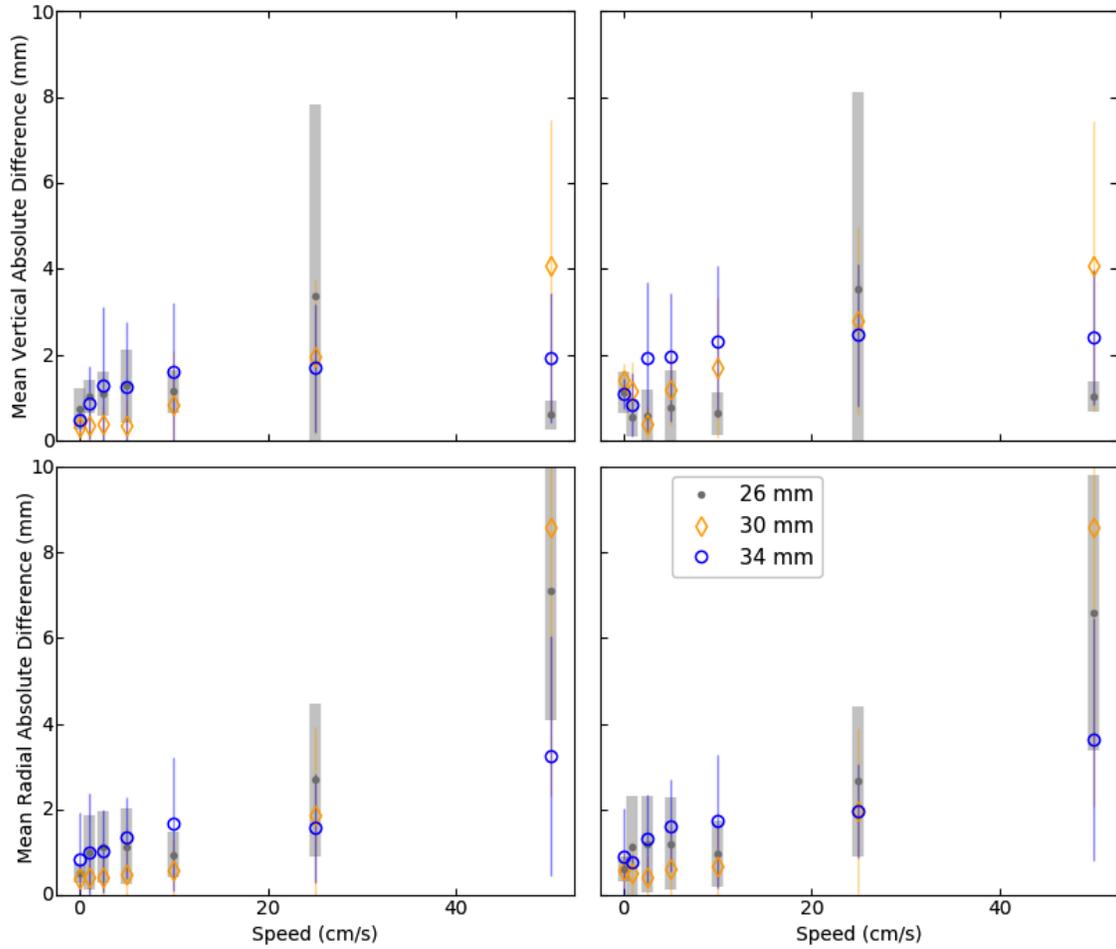

*Figure 6* – The differences between the ring reconstruction using the check cable, the manufacturer measured source path (left panels) and the clinically extracted path (right panels) as a function of the check cable speed. (repeated from page 9)

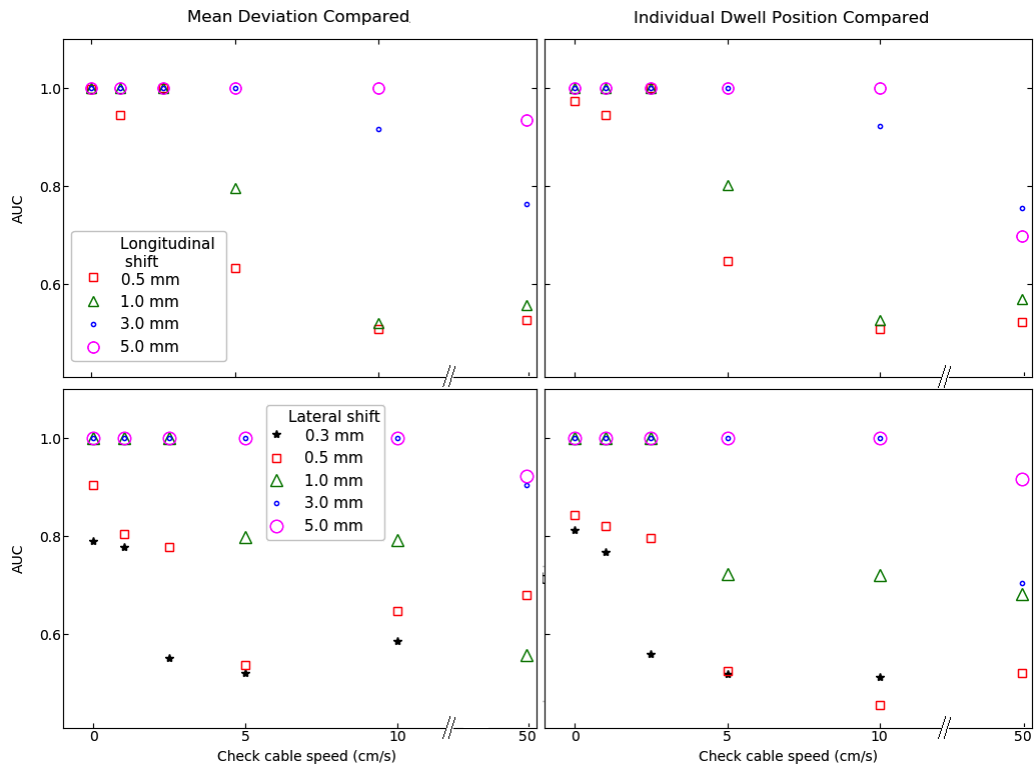

*Figure 7* – Area under a receiver operator curve as a function of the check cable speed for known indexer's length (upper panels) and lateral (lower panels) shifts. Both error criteria were tested: the right panels use the mean deviation while the left panels use individual dwell positions with a threshold at 50 % as presented in the method section. (repeated from page 10)